\documentclass[10pt]{article}
\usepackage{graphicx}

\usepackage{epsfig}
\usepackage{latexsym}
\usepackage{amsmath}

\begin{document}

\title{\Large \bf Assessing the true role of  coauthors in the   $h$-index measure of an author scientific  impact  }

\author{ \large \bf  M. Ausloos$^{1,2,}$\footnote {marcel.ausloos@ulg.ac.be,}  
\\ $^1$ e-Humanities Group, KNAW,\\Joan Muyskenweg 25, 1096 CJ Amsterdam, The Netherlands  \\$^2$   GRAPES, rue de la Belle Jardiniere, B-4031 Liege, \\Federation Wallonie-Bruxelles, Belgium
 }

\maketitle

  \begin{abstract}
A method based on the classical principal component analysis leads to demonstrate that the role of co-authors should give a  $h$-index measure to a group leader \underline{higher} than usually accepted.  
The method rather easily gives what is usually searched for, i.e. an estimate of the role (or "weight") of co-authors, as the additional value to an author papers' popularity.  The construction of the  co-authorship popularity $ \mathcal{H}$-matrix is exemplified and the role of eigenvalues and the main eigenvector component are discussed. An example illustrates the points and serves as the basis for suggesting  a generally practical application of the concept.
\end{abstract}

 PACS:  ...
 
 \textit{Keywords:}  ...
 \maketitle
 
  \section{Introduction}
  
The $h$-index value of an author results from the counting of his/her quoted publications   \cite{hindex}, ranked according to their popularity (the most quoted paper gets a  rank $r$=1, etc.), and is obtained  by the rank value  ($h=r$) such that the papers above that rank ($r\ge h$) have less citations than $h$.   The $h$-index has been invented to quantify an author impact, though it is rather a measure of an author paper productivity and/or popularity \cite{braun85,beck},- which maybe partially due to some paper content quality  \cite{Radiology255.10.342 bibliom} or to co-author fame \cite{MABSS}.  
   
   It has been much discussed what publication  is (or has to be) considered, when measuring $h$. Sometimes book citations are not counted; sometimes,  there is double counting, or sometimes two papers deposited on different websites are counted as different papers, or sometimes not;  sometimes papers in proceedings are not (or sometimes are) counted  as of equal value as those in classical peer-review journals, etc \cite{Buchanan,Vanclay,AlonsoetJOI3.09}.  Therefore, $h$ depends on the type of  selection criteria and big data search engine: Google Scholar (Google), Web of Science (Thomson Reuters),  Scopus (Elsevier) databases \cite{[BIJ]}, etc.  However, the following considerations apply to all search engine results.

   Many variants  have been proposed in order to remediate  several so called defects   \cite{SCIM85.10.741hbar}-\cite{JASIST59.08.830BornmannMutzDaniel} of the original $h$-measurement\footnote{see also  http://sci2s.ugr.es/hindex/ }. Sometimes self-citations have been scorned upon \cite{Schreiber2007EPL}, sometimes not  \cite{Ausloos08}. Another, considered very important, criticism has been  about the counting of coauthors and their role  \cite{MelinPersson96}-\cite{[GR]}.   It has been often argued that the  number of quotations  of a paper should be weighted according to the number of coauthors of this paper, thereby reducing the  $h$-index of an author having many co-authors, as in the  $ Ôprofit (p)-indexÕ $ \cite{profitindexPLOS}. It is true that sometimes team leaders are unaware of the papers they have published. Sometimes there is complaisance co-authorship as well   \cite{Kwok05}.	 
 
  
   The present paper has for  main aim to  propose a practical and basically sound (i.e. physics-prone)  way  of remeasuring the $h$-index (keeping the same  "name" ($h$)  and quasi similar notations as in the "$h$-index" literature, for simplicity)  of an author publishing with co-authors.  It is argued that the original $h$-index, in fact, undervalues the role of co-authors.  The following study and  method, therefore, emphasize  that the impact  of a team leader, or more generally co-authorship,  is underrepresented by the classical $h$-index.
   
   The "theory arguments" seem  to follow better from practical examples, in a deductive way rather than through an inductive presentation.  The methodology idea is based on the principal component analysis (PCA) method which aims  at  reducing the dimensionality of a data set, consisting of a large number of interrelated variables, while retaining as much as possible of the variation present in the data set. 
This is achieved by transforming the raw data into a new set of measures,
the principal components (PCs), which are uncorrelated, and which are ordered so that the  first few retain most of the variations present in all of the original variables. Here, the data set is the $h$-index values for authors, but considering that they can have 1, 2, 3, ... co-authors, forming teams. For these teams, one can calculate  also the corresponding $h$-index, in a usual way. This leads to write down a square matrix with dimensions equal to the number of considered co-authors. To calculate the eigenvalues and eigenvectors is next  a classical matter. Then, the result leads the true measure of co-authorship  popularity from the  set of ranked papers of such co-authors. The example cases, illustrating the argument,  are  limited to a few co-authors, but could obviously be extended.  Their finite size is wholly irrelevant; in fact,this allows a better comprehension of details. 
Other extensions are briefly discussed in the conclusion section.
 
 \section{Methodology}\label{methodology}
  
 At the start, get the $h$-index  ($h_{ii}$) for each authors ($i$),  from some search engine, e.g., Google Scholar or Web of Science. The source is irrelevant, since it will be seen that the method applies whatever the search engine.  Of course,  initial and final numbers will be different, but the discussion on whether some source is "better"  than another is not part of the present development. 
 
 Next,  reduce  the publication list  of the $i$ authors only to the  joint papers  by  a couple of co-authors, e.g. $i$ and $j$, i.e. $N_{i,j}$.  Thus,  after ranking these papers, one easily obtains   the equivalent of the $h$-index, i.e. an $h_{ij}$,  for the  couple of authors, from the list of  $N_{i,j}$  papers.  A warning: this list might include papers   which have a "large" citation number, yet not large enough to have a rank lower than $h_{ii}$  for author $i$ (or $j$). 
  Indeed, for the author $i$ a paper might not be often quoted, whence have a rank higher than  $h_{ii}$,  although   such a number of citations might be important enough to have a rank lower than  the $h$-index for the "couple of authors", i.e. $h_{ij}$.

 There is no need to emphasize that  the citation lists  should be taken from the same data base,  for coherence purpose. A practical point  can be also mentioned. It is useful to cross-check the lists, i. e.  repeating the procedure stating from $j$, and obtaining $h_{ji}$, thereby   observing that one truly obtains  $h_{ij}\equiv  h_{ji}$.

Thereafter, define the  co-authorship popularity $ \mathcal{H}$-matrix, having $h_{ij}$ as its off-diagonal elements and has  $h_{ii}$ on the diagonal. The order $i$ is irrelevant. However, for  a discussion,  it   seems appropriate to rank the authors $i$ according to their $h_{ii}$ value. In so doing,  $h_{ij}$ (or $h_{ji}$ ) $ \le  h_{jj}  \le  h_{ii}$.
 This matrix   $\mathcal{H}$ differs from the co-occurrence matrices introduced in  \cite{Morris2005, JASIST48.07.1764coauth} which only consider the frequency of partnerships. 
 
 Finally, calculate the eigenvalues and eigenvectors of  $ \mathcal{H}$. For emphasizing the partner weights, the lowest component of the eigenvector   corresponding to the largest eigenvalue has always been   imposed  equal to 1.
 
  \section{A real case}\label{acase}
  
  Consider the following ($i= 1, 2, 3, 4)$ co-authors :
MAU, PCL, APE,  and JPE, respectively, having worked in statistical mechanics  independently, together,  or with various co-authors. A few characteristics of their publication lists is given in Table \ref{MAUPCLAPEJPEstat}.
Next, take the whole publication list of each author, e.g. from Google Scholar, without any loss of generality for the argument.

For the present case, one obtains (six) 
2x2 matrices; the same procedure is repeated for finding the matrix elements of the (four) 3x3 matrices, and  for the unique   4x4 matrix.   The number of joint papers is of course \underline{not} increasing in this process.  Recall that it seems convenient to order the authors   ($i= 1, 2, 3, 4$) according to their $h$-index.

For space saving, the  $ \mathcal{H}$-matrices of this example are displayed below,   with on the same line, the relevant number of joint papers;    the matrix eigenvalues, but  only the (unnormalized) eigenvector  components corresponding to the largest eigenvalue (designated by $^{(1)}$)  are  also  given here below. For immediately emphasizing the partner weights, the lowest component of this eigenvector is imposed to be equal to 1; also the index $_i$ of the component refers to the \underline{author} rather than  to its  usual order when writing a vector.

 The  6 matrices emphasizing  links between two authors, among the 4 considered,  are 
\begin{eqnarray*}  \label{MAPCe12} \nonumber
\phantom{}h_{MAU,PCL}  =\left( \begin{tabular}{ll} 35 & $10$  \\  $10$ & $11$\end{tabular} \right);\;\;\ \;N_{1,2}= 30;  
 \\ =>  \lambda^{(1)}_{1,2}= 38.620; \;\; \lambda^{(2)}_{1,2} = 7.380 ;  \;\;\; \;\;x_1^{(1)}=2.765  \;\;\; \;\;x_2^{(1)}=1
   \end{eqnarray*} 
\begin{eqnarray*}  \label{MAAPe13} \nonumber
\nonumber  \nonumber     \\h_{MAU,APE}  =\left( \begin{tabular}{ll} 35 & $\;\;7$  \\  $\;7$ & $10$\end{tabular} \right);\;\;\; \;\;N_{1,3}=21;
\\Ê=>  \lambda^{(1)}_{1,3}= 36.827; \;\; \lambda^{(2)}_{1,3} = 8.173;  \;\;\; \;\;x_1^{(1)}=3.841 \;\;\; \;\;x_3^{(1)}=1
\end{eqnarray*} 
\begin{eqnarray*}  \label{MAJPe14}\nonumber
      \\h_{MAU,JPE}   = \left( \begin{tabular}{ll} 35 & $\;\;2$  \\  $\;\;2$ & $\;\;2$\end{tabular} \right);\;\;\;N_{1,4}=2; \;\; 
\\  =>\lambda^{(1)}_{1,4}=  35.121 \;\; \lambda^{(2)}_{1,4} =  1.879;  \;\;\; \;\;x_1^{(1)}=16.63   \;\;\; \;\;x_4^{(1)}=1
\end{eqnarray*} 
\begin{eqnarray*}  \label{PCAPe23}\nonumber
\phantom{} \nonumber     \\h_{PCL,APE}  =\left( \begin{tabular}{ll} 11 & $\;\;6$  \\  $\;\;6$ & $10$\end{tabular} \right);\;\;\;N_{2,3}=8; \;\; 
 \\Ê=>  \lambda^{(1)}_{2,3}= 16.521 ; \;\; \lambda^{(2)}_{2,3} =  4.479;  \;\;\; \;\;x_2^{(1)}=1.087  \;\;\; \;\;x_3^{(1)}=1
\end{eqnarray*} 
\begin{eqnarray*}  \label{PCJPe24}
 \nonumber     \\h_{PCL,JPE}  =\left( \begin{tabular}{ll} 11 & $\;\;2$  \\  $\;\;2$ & $\;\;2$\end{tabular} \right);\;\;\; \;N_{2,4}=2; \;\; 
\\   =>  \lambda^{(1)}_{2,4}= 11.424  ; \;\; \lambda^{(2)}_{2,4} =   1.576;  \;\;\; \;\;x_2^{(1)}=4.702  \;\;\; \;\;x_4^{(1)}=1
\end{eqnarray*} 
\begin{eqnarray*}  \label{APJPe34}
 \nonumber     \\h_{APE,JPE}  =\left( \begin{tabular}{ll} 10 & $\;\;2$  \\  $  \;\;2$  & $\;\;2$\end{tabular} \right);\;\;\; \;N_{3,4}=2;\;\; 
\\Ê  =>  \lambda^{(1)}_{3,4}=10.472  ; \;\; \lambda^{(2)}_{3,4} = 1.528  ;  \;\;\; \;\;x_3^{(1)}=4.230  \;\;\; \;\;x_4^{(1)}=1
\end{eqnarray*}

 The  4 matrices emphasizing the  links between three authors, among the 4 considered,  are 
\begin{eqnarray*}  \label{MAPCAPe123gr}
\phantom{}    h_{1,2,3}  =\left( \begin{tabular}{lll} 35& $10$ &$\;\;7$\\  $10$ & $11$&$\;\;6$\\ $  \;\;7$&$  \;\;6$&10 \end{tabular} \right);\; \;N_{1,2,3}=8;\;\;  \\
=>  \lambda^{(1)}_{1,2,3}= 41.041 ; \;\; \lambda^{(2)}_{1,2,3} =  10.620; \;\; \lambda^{(3)}_{1,2,3} =  4.338;
\\ 
  \;\;\;    \;\; x_1^{(1)} = 3.318;\;\;\; x_2^{(1)}=1.304;\;\;\;x_3^{(1)}=1 
\end{eqnarray*} 
 \begin{eqnarray*}  \label{MAPCJPe124}
  \nonumber      h_{1,2,4}=\left( \begin{tabular}{lll} 35& $10$ &$\;\;2$\\  $10$ & $11$&$\;\;2$\\$\;\;2$&$\;\;2$&$\;\;2$
\end{tabular} \right);\; \;N_{1,2,4}=2;\;\; \\
=> \lambda^{(1)}_{1,2,4}=38.799  ; \;\; \lambda^{(2)}_{1,2,4} =7.626; \;\; \lambda^{(3)}_{1,2,4} =1.575;  
 \\ 
  \;\;\;    \;\;x_1^{(1)} = 13.388;\;\;\; x_2^{(1)}=4.888;\;\;\;x_4^{(1)}=1 
 \end{eqnarray*} 
\begin{eqnarray*}   \label{MAAPJPe134}
  \nonumber      h_{1,3,4}=\left( \begin{tabular}{lll} 35& $\;\;7$ &$\;\;2$\\  $\;\;7$ & $ 11 $&$\;\;2$\\$\;\;2$&$\;\;2$&$\;\;2$
\end{tabular} \right);\; \;N_{1,3,4}=2;\;\; \\ => \lambda^{(1)}_{1,3,4}=37.064 ; \;\; \lambda^{(2)}_{1,3,4} = 9.369 ; \;\; \lambda^{(3)}_{1,3,4} =  1.567;
 \\\;\;\;    \;\; x_1^{(1)} = 13.743;\;\;\; x_3^{(1)}=3.771;\;\;\;x_4^{(1)}=1 
 \end{eqnarray*} 
\begin{eqnarray*}  \label{PCAPJPe234a}
\nonumber    h_{2,3,4}  =\left( \begin{tabular}{lll} 11& $\;\;6$ &$\;\;2$\\  $\;\;6$ & $10$&$\;\;2$\\$\;\;2$&$\;\;2$&$\;\;2$\end{tabular} \right);\; \;N_{2,3,4}=2;\;\; \\
=>  \lambda^{(1)}_{2,3,4}=  17.051; \;\; \lambda^{(2)}_{2,3,4} =4.484  ; \;\; \lambda^{(3)}_{2,3,4} = 1.465;
\\ 
  \;\;\;  \;\; x_2^{(1)} = 3.903;\;\;\; x_3^{(1)}= 3.605 ;\;\;\;x_4^{(1)}=1.
\end{eqnarray*} 

The  matrix  emphasizing  the links between the  four authors is 
\begin{eqnarray*}  \label{4x4}
\nonumber  h_{1,2,3,4}=\left( \begin{tabular}{llll} 35& $10$ &$\;\;7$&$\;\;2$\\  $10$ & $11$&$\;\;6$&$\;\;2$\\$\;\;7$&$\;\;6$&10&$\;\;2$\\ $\;\;2$&$\;\;2$&$\;\;2$&$\;\;2$
\end{tabular} \right); \;\;N_{1,2,3,4}= 2; \; \\ => \lambda^{(1)}_{1,2,3,4}=41.277  ; \;\; \lambda^{(2)}_{1,2,3,4} =10.921  ; \;\; \lambda^{(3)}_{1,2,3,4} =  4.339  ; \;\; \lambda^{(4)}_{1,2,3,4} =  1.463;
\\%
  x_1^{(1)} = 11.539;\ \; x_2^{(1)} = 4.576;\;\; x_3^{(1)}= 3.524;\;\;x_4^{(1)}=1. \nonumber \\
\end{eqnarray*} 

N.B. Those 4 authors  have only 2 papers in common.

\begin{table} \begin{center} 
\begin{tabular}[t]{lcccc} 
\\ 
\hline 
$co-authors$ $i $: &\phantom{fBm}MAU&\phantom{fBm}PCL  & \phantom{fBm}APE& \phantom{fBm}JPE
\\ 
\hline 
\phantom{fBm} $h_{ii}$ &\phantom{fBm}35&\phantom{fBm} 11&\phantom{fBm} 10&\phantom{fBm} 2\\
N. citations of most cited paper&\phantom{fBm}152&\phantom{fBm}127&\phantom{fBm} 37& \phantom{fBm} 7\\ 

N. citations till $h$&\phantom{fBm}1113&  \phantom{fBm} 296 &\phantom{fBm}224&\phantom{fBm} 14\\
N. coauthors &\phantom{fBm}317&  \phantom{fBm} 32&\phantom{fBm} 46& \phantom{fBm} 4\\ 
N. papers with $"best"$ coauthor &\phantom{fBm}155&  \phantom{fBm} 30&\phantom{fBm} 21& \phantom{fBm} 2\\ 
N. publications ($<$2012) &\phantom{fBm} 571&  \phantom{fBm} 34&\phantom{fBm} 111&\phantom{fBm}2\\ 
\hline
\end{tabular} 
   \caption{Productivity characteristics of the 4 co-authors considered in the text}\label{MAUPCLAPEJPEstat}
\end{center} \end{table}

     \section{Case analysis and implications}\label{caseanalysis}
It can be immediately observed that  
the  (here called) "average $h$-index" for MAU, resulting from having co-authored papers at least with  PCL or with APE or with JPE,  leads to a 
$<h>_2^{(1)}$ = (38.62 + 38.83 + 35.12 ) / 3 = 37.52,
instead of   $<h>_ 1^{(1)}( \equiv h_{11})$ = 35. 

 

In the same line of thought,  consider the  (average) $h$-index for MAU resulting from having co-authored papers   at least  
   with  PCL and  with  APE
or  MAU with  PCL  and  with  JPE
or  MAU with  APE  and  with  JPE. It  easily  found  that
$<h>_3^{(1)}$ = (41.04 + 39.80 + 37.06 ) / 3 = 39.30.        

  The effective $h$-index value  can be calculated, in a similar manner, for another author, e.g. APE,  due to his partnership in  this particular 4-member  team. It is easily obtained that 
  $<h>_3^{(2)}$ = (17.05 +10.62+7.626)/3= 11.77, instead of  APE   $h_{22}$= 11.
 
   Thus, one can  emphasize that $  <h>_4^{(1)} \equiv \lambda^{(1)}_{1,2,3,4}  $=  41.277  is not some "average", but is the truly effective value for MAU  due to publishing (and being quoted) when participating in this  group of 4 co-authors. 
    
    Moreover,  the  largest  principal component is also giving some relevant information on the relative impact of a co-author. It is sufficient to normalize the vector components indeed and consider the absolute weights.  For example, for MAU in the 4-member team,   the largest PC  is  found equal to  11.539/ $\sqrt (11.539^2+4.576^2+3.524^2+1)$ $\sim$  89\%. This results in the effective $h$ due to team partnership being equal to  (41.277 x 0.89 $\simeq$) 36.80; in contrast to the raw value 35 which is not taking into account various co-authorships.
         
         The output due to the eigenvector components as  indicators  and measures  of the respective weight gains and losses is postposed to Sect.  \ref{caseShortreal} for better emphasis of the method interest.
         
    Note that the argument on the proportionality factor can be applied to each level or participation, considering sub-groups of co-authors.  
         
         \section{Two other cases}\label{2cases}
         A reviewer of the initial version of this paper claimed {\it  that the conclusion is based on the use of a very small sample of 4 authors, which in addition is not arbitrary since it includes the author himself. So contrary to the author claim that he uses "an arbitrarily selected example, but without loss of generality", he is using a specific sample not at all arbitrarly}  (sic).  {\it At least a few more cases should be treated to provide a more solid basis to the idea.
}

Therefore, two other cases are outlined, though without going through the complete details as above.

\subsection{Extended "real case"}\label{caseExtreal}
         
         Consider  two other authors, ($i=5$) JMK and   ($i=6$)  DAH, having worked  with the previous 4 authors, but not all of them. Therefore  a few  off-diagonal  elements  are necessarily vanishing.  N.B. Those 6 authors  have  no  paper  in common: $\;\;N_{1,\dots,6}= 0.$ Moreover, for comparison with the above, the authors have not been ranked according to their $h-$index when writing
     the  matrix  emphasizing  the links between the  six authors  
\begin{eqnarray*}  \label{6x6}
\nonumber  h_{1,\dots,6}=\left( \begin{tabular}{llllll} 
$35$& $10$ &$\;\;7$&$\;\;2$&$\;\;3$&$\;\;2$\\  
$10$ & $11$&$\;\;6$&$\;\;2$&$\;\;3$&$\;\;2$\\ 
$\;\;7$&$\;\;6$&$10$&$\;\;2$&$\;\;3$&$\;\;0$\\ 
$\;\;2$&$\;\;2$&$\;\;2$&$\;\;2$&$\;\;1$&$\;\;0$\\ 
$\;\;3$&$\;\;3$&$\;\;3$&$\;\;1$&$\;\;9$&$\;\;2$\\ 
$\;\;2$&$\;\;2$&$\;\;0$&$\;\;0$&$\;\;2$&$17$\\ 
\end{tabular} \right)  ;
\\ \\ =>  \lambda^{(1)}_{1,\dots,6}=42.232  ; \;\; \lambda^{(2)}_{1,\dots,6} =17.207 ; \;\; \lambda^{(3)}_{1,\dots,6} =  12.262 ;\\Ê\\  \;\; \lambda^{(4)}_{1,\dots,6} =  6.658;
\;\; \lambda^{(5)}_{1,\dots,6} =  4.188;
\;\; \lambda^{(6)}_{1,\dots,6} =  1.452,
\\%
 \nonumber \\
\end{eqnarray*}
with
\begin{eqnarray*}  \label{6x6ev}
\nonumber  \;\;  x_1^{(1)} = 11.167;\ \; x_2^{(1)} = 4.577;\;\; x_3^{(1)}= 3.513;
\\ \\   \; x_4^{(1)} = 1.0;\;\; x_5^{(1)}= 1.859; \;\;x_6^{(1)}=1.397. \nonumber \\
\end{eqnarray*}  

 It is noticed that the main author weight goes down from 11.539 to 11.167. This is due to the influence of the sixth author which has a  $h_{6,6}$=17, in contrast to the 2nd author who has  a smaller $h_{2,2}$= 11.  Therefore the method allows to test the loss (or gain, in the previous section) of some  author's influence  due to some    co-author.
         
         \subsection{Shortened "real case"}\label{caseShortreal}
         
        Indications on the weight gains and losses are best described on a "shortened case", i.e. when the number of components of the eigenvectors, or the matrix rank, is small.  In contrast to a 6 member team, consider the real case with  three (for concision)  authors like MD, SG and AP who all have a rather high $h_{i,i}$ (as of Oct. 2014) such that the popularity matrix reads
 \begin{eqnarray*}  \label{MDSGAPEV123}
 h_{MD,SG,AP}  =\left( \begin{tabular}{lll} 35& $\;\;3$ &$\;\;8$\\  $\;\;3$ & $30$&$\;\;0$\\ $  \;\;8$&$  \;\;0$&12 \end{tabular} \right);
\\Ê\\Ê=>  \lambda^{(1)}_{MD,SG,AP}= 38.479 ; \;\; \lambda^{(2)}_{MD,SG,AP} =  29.070; \;\; \lambda^{(3)}_{MD,SG,AP} =  9.451;
\\ 
\end{eqnarray*} 
         
       In this (real) case SG and AP have no paper in common. However, both other couples (MD, SG) and (MD,AP) have a few papers in common,  although hey are often cited, whence the relatively small   $h_{12}$ and $h_{13}$.  It is of interest to write the three   (normalized) eigenvectors:
   \begin{eqnarray*}  \label{MDSGAPev123}
 x ^{(1)} =  ( 0.303;-0.907;0.293);\\
 \;\;\; x^{(2)}= (-0.044;-0.321;0.946);\\
 \;\;\;x ^{(3)}= (-0.952;-0.274;0.137).
\end{eqnarray*} 

Two points can be made as a brief conclusion of this subsection. On one hand, as before, the "highest $h-$" author, i.e., $i=1$ gains from the other two,  $h_{11} =  35 \rightarrow \simeq 38.5$, but the second (or middle one, in this case) does not loose much  ($h_{22} =  30 \rightarrow \simeq 29$) in having no publication, whence  of course no citation, with the third partner. On the other hand, it is well seen that the eigenvectors indicate and measure  the respective weight gains and losses.

   \section{Conclusions}\label{conclusions}
   
   In  summary, the above indicates that the effect of co-authors  on evaluating the popularity of an author through the $h$-index method  can be investigated through a principal component analysis method.   Through an arbitrarily selected example,  but without loss of generality,  it has been proved that the $h$-index undervalues the role of  the team,  in particular on the team "leader".  Two other cases have served to indicate that the method can be applied in larger or smaller samples. It is found that  an effective $h$-index can be calculated from the  co-authorship popularity $ \mathcal{H}$-matrix  eigenvalues, through the selection of team partners, but also up to the whole team  size. 

It has been remarked that the co-authorship popularity $ \mathcal{H}$-matrix is sensitive to the size of the team and to the own $h$-index of the various members, - as should be expected, but also on the joint $h$-index of co-author couples.   The relative weight is therefore nicely measured when imagining team members influence, e.g. on a finished project. There would be no need for {\it a posteriori}  (and previously rather non objectively) asking the weight of a co-author, as it is often done in some dossier evaluation (see also comments on such a consideration in the Appendix).

    An interesting application would  occur when  the ranking of teams with a leader is necessary in  fund raising processes, but also in hiring and promotion processes when the team partnership capacity of an author has to be quantified.  
    
    The above demonstration seems  easily applied through any web search engine, if, e.g., evaluation committees wish to consider a specific search engine.  
   
 The approach seems much more realistic and valuable than the classical $h$-index  and several variants, - the more so nowadays in presence of co-authorship inflation.  The method, based on a standard physics approach,  is fundamentally different from   Schreiber's $h_m$ \cite{Schreiber2008c} and the many   variant  (fractional \cite{galamscientom12}, normalized \cite{sidiropoulosscientom10}, ...) estimates of the $h$-index (see Appendix).
 
In that respect, a final note  about the construction of the  co-authorship popularity $ \mathcal{H}$-matrix seems of interest. The usual  co-authorship network considerations suggest that one could examine a not-necessarily symmetric matrix, but evaluate the $h_{ij}$ elements, taking into account the order of authors, i.e. $h_{ij} \neq h_{ji}$. This would not change much the eigenvalues, but would modify the eigenvectors and the relative  weights.  However,  it is well known that the order of authors obeys different criteria, with different justifications,  in different scientific fields   \cite{JoI7.13.198AbramoorderCA}.  In the main text, the position of  the co-author in the list has not been a criterion. Nevertheless, this consideration  could be easily  implemented.
       
  In conclusion, it can be claimed that the  PCA method  rather easily gives what is searched for, i.e. an estimate of the role of co-authors, as \underline{the} additional  weighted value to the measure of an author research paper's popularity within the $h-$index scheme.
 
\vskip0.8cm
 \begin{flushleft}
{\large \bf Acknowledgment}
\end{flushleft}
Thanks to the editors   for inspiring me into writing this work. I gratefully acknowledge  stimulating discussions with many wonderful colleagues at  several meetings of the ESF Action
COST MP-0801 'Physics of Competition and Conflict'.   I thank all colleagues  mentioned in the text and bibliography for  providing relevant data. This paper is part of scientific activities in COST Action TD1210 'Analyzing the dynamics of information and knowledge landscapes'.

\vskip0.8cm
  \begin{flushleft}
{\large \bf Appendix}
\end{flushleft}
This Appendix follows a remark by a reviewer:
{\it In addition, the author claims that "The method, based on a standard physics approach, is fundamentally different from   Schreiber's $h_m$ \cite{Schreiber2008c} and the many 
 variant  (fractional \cite{galamscientom12}, normalized \cite{sidiropoulosscientom10}, ...) estimates of the h-index." without proving it. Indeed, he should compare those different methods using the same sample.}
 
 The point is of interest, but the reviewer is misunderstanding the  main point. There is no  need for a long proof  to show that something is obviously different:  to any reader, the present method should appear to be (completely) different from these weighting the number of citations through the number of authors, in the cited references. The latter methods {\it in fine}  lead to a decrease of the apparent  $h$-index of the "team leader", while the present approach shows that   its value should be considered to be increased instead, (because co-authors  unduly take some part of the popularity value of the team leader)!
 
 Nevertheless, discussion of the fractional aspect being much related to the present work, values of the  fractional $h_{ii}$ index,  according to \cite{Schreiber2008c} and  \cite{galamscientom12}  can be calculated as an illustration.  However, the comparison with all the normalized   indices introduced in  \cite{sidiropoulosscientom10}   or others discussed by Schreiber \cite{Schreiber2010b}  is not  going to be made here in order to keep the size of the appendix at a reasonable length. On the other hand, it  seems  sufficient, for the present point, only  to consider the data of the Sect. \ref{caseShortreal} case.

In summary,  Schreiber  \cite{Schreiber2008c}   proposes to count each paper  citation only fractionally according to  the inverse of the number of authors.  In doing so, the following results   $h_{11}$ = 22, $h_{22}$ = 25, $h_{33}$ = 5 are obtained  for the Sect. \ref{caseShortreal} case.   Note here that SG  ($i=2$) has rarely more than one co-author and is much publishing single author papers, whence his fractional $h$-index remains very high.  
Taking the off-diagonal element fractionalized as well into account, one reaches
$\lambda^{(1)}_{MD,SG,AP}= 27.206;   \lambda^{(2)}_{MD,SG,AP} =  21.185;   \lambda^{(3)}_{MD,SG,AP} =  3.609$, all obviously smaller than in the non-fractionalized case, but also showing  a similar evolution toward an increase of the "leader value" with respect to the other authors.

In a more complicated way, Galam's tailor based allocation  (TBA)   \cite{galamscientom12}  gives a set of  weight possibilities, - although without providing an optimum one.  The weight of an author is (in short) related to his/her position in the co-author list, but is (expectedly and supposedly) decided by the co-authors. In order to get a finite size appendix here, a specific constraint has to be  selected for finding the fractionalized $h_{ij}$.  In order to contrast with the uniform distribution in   Schreiber's approach,  a practically admitted constraint can be implemented for this illustration. The choice  of   the weight given to an author at position $p$ of a given $q$ authors paper, i.e., $g(p,q)$, is hereby made similar to that used   in  evaluating rules  at FNRS (Fonds National de la Recherche Scientifique) in Belgium. Let the value of a paper be $2q$. For a paper with two authors, each one gets the same weight ($1/q=$50\%). Otherwise, the first author gets 50\% of the weight; the last author gets  25\%, the rest being  equally divided between the other authors.  It is obvious that the weight of the "middle list" co-authors gets quickly small  when their  number increases. Implementing such a rule leads in the Sect. \ref{caseShortreal} case to 
 $h_{11}$ = 20, $h_{22}$ = 24, $h_{33}$ = 6.
 Taking the off-diagonal element, fractionalized as well, into account, one reaches
$\lambda^{(1)}_{MD,SG,AP}= 24.339;   \lambda^{(2)}_{MD,SG,AP} =  18.300;  \lambda^{(3)}_{MD,SG,AP} =  5.362$. Again, observe the  strong effect of publishing with only a few co-authors and taking the first or last place in the list. 

It is worth as a conclusion to quote from \cite{galamscientom12}: {\it  the TBA rescaling disadvantages senior authors who usually sit last with several co-authors ....  (while) ...
a low citation paper does contribute mainly to first and last authors.} It could be added that according to the here  called FNRS rule or  constraint   implemented as in the TBA, it is not  very interesting to be one among many co-authors, - the more so if the paper is not often cited, and if the position in the list is toward the middle.  Fortunately, the present approach corrects a little bit the drastic rule. Nevertheless, this indicates that  the {\it a priori} choice of the $g(p,q)$ weight also much  influences the final values, as expected, - but the index evolution is again the same: the "leader" gets  a higher  $h_{ii}$ value to the detriment of his/her colleagues. 

Notice that as in \cite{Schreiber2008c},  the present method {\it   does not need a rearrangement of the citation records} in contrast to the fractional methods,   and {\it ... it is not sensitive to extreme values of the number of co-authors ...  cannot decrease when the number of citations increases, and ...  its construction does not push highly cited papers out of the core}. 

In conclusion, it seems obvious that these approaches (\cite{Schreiber2008c}, \cite{galamscientom12},  \cite{sidiropoulosscientom10}, ...)  giving an {\it a priori}  weight according to the position in a co-author list are different from the present one.
. 

 \clearpage

\begin{thebibliography}{99}

  \bibitem{hindex}  J. E. Hirsch,  An index to quantify an individual's  scientific research output, Proc. Nat. Acad. Sci. USA 102  (2005) 16569-16572.  
  
\bibitem{braun85}
T. Braun, W. Gl{\"a}nzel, A. Schubert,  Scientometric indicators. A 32-country comparative  evaluation of publishing performance and citation impact. (World Scientific, Singapore, 1985).
			
\bibitem{beck} 
I. M. Beck, A method of measurement of scientific production, Science of Science   4  (1984) 183-195.

 \bibitem{HESPRE2010career} 
 A.M. Petersen, F. Wang,   H.E. Stanley, Methods for measuring the citations and productivity of scientists across time and discipline, Phys. Rev. E  81 (2010) 036114.
 
    \bibitem{Radiology255.10.342 bibliom}    V. Durieux, P.A.  Gevenois,   Bibliometric Indicators:
Quality Measurements of Scientific Publication,   Radiology 255   (2010) 342-351

    \bibitem{MABSS}  M. Ausloos,  Binary Scientific Star Coauthors Core Size,  Scientometrics  99 (2014) 331-351.
  
\bibitem{Buchanan}  R. A. Buchanan. Accuracy of cited references: The role of citation databases,  College and Research Libraries 67  (2006) 292-303.

 \bibitem{Vanclay} J. K. Vanclay,  On the robustness of the h-index,  J.    Am. Soc. Inf. Sci. Technol. 58  (2007)  1547-1550.  
 
\bibitem{AlonsoetJOI3.09}
S. Alonso, F.J. Cabrerizo, E. Herrera-Viedma, F. Herrera,  h-Index: A review focused in its variants, computation and standardization for different scientific fields,  J. Inform. 3 (2009) 273-289.

   \bibitem{[BIJ]} 
J.  Bar-Ilan,  Which h-index? - A comparison of WoS, Scopus and Google Scholar, Scientometrics,  74 (2008)  257-271.

  \bibitem{SCIM85.10.741hbar} J.E. Hirsch, An index to quantify an individualÕs scientific research output that takes into account the effect of multiple coauthorship,  Scientometrics   85 (2010) 741-754.
  
  \bibitem{Schreiber2010b} M. Schreiber,  Twenty Hirsch index variants and other indicators giving more or less preference to highly cited papers. Ann. Phys. Berlin) 522  (2010) 536-554. 
   
\bibitem{Schreiber2012JoI6_17v}    M. Schreiber, C.C.   Malesios, S.  Psarakis, Exploratory factor analysis for the Hirsch index, 17 h-type variants, and some traditional bibliometric indicators, J. Inform.  6   (2012)  347-358. 

 \bibitem{JASIST59.08.830BornmannMutzDaniel} L. Bornmann, R.  Mutz,  H. Daniel,  Are there better indices for evaluation purposes than the h-index? A comparison of nine different variants of
the h-index using data from biomedicine, J.    Am. Soc. Inf. Sci. Technol.   59 (2008) 830-837.

\bibitem{Schreiber2007EPL} M. Schreiber,   Self-citation corrections for the Hirsch index,  Europhys. Lett.  7  (2007)  30002. 

 \bibitem{Ausloos08}  M. Ausloos, R. Lambiotte, A.  Scharnhorst, I. Hellsten,  Andrzej Pekalski networks of scientific interests with internal degrees of freedom through self-citation analysis,   Int. J. Mod. Phys. C  19    (2008)  371-384.  

  \bibitem{MelinPersson96} G. Melin,  O. Persson, Studying research collaboration using co-authorships, Scientometrics  36   (1996)  363-377.
  
  \bibitem{KBLDWMKAV05}   K. B\" orner, L.  Dall'Asta, W.  Ke, A.  Vespignani, Studying the emerging global brain: Analyzing and visualizing the impact of co-authorship teams,   Complexity  10   (2005) 57-67.
 
 \bibitem{SCIM51.01.69Glanzel} W. Glanzel, Coauthorship Patterns and Trends in the Sciences (1980-1998): A Bibliometric Study with Implications for Database Indexing and Search Strategies,  Scientometrics  51 (2001) 69-115, 

 \bibitem{Laudel} G. Laudel, What do we measure by co-authorships? In M. Davis \& C. S. Wilson (Eds.) Proceedings of the 8th International Conference on Scientometrics and Informetrics (pp. 369-384). Sydney, Australia: Bibliometrics \& Informetrics Research Group  (2001).   

  \bibitem{SCIM68.06.179hav} P. D. Batista, M. G. Campiteli, O. Kinouchi, A. S. Martinez, 
 Is it possible to compare researchers with different scientific interests?
Scientometrics  68  (2006)  179-189. 

 \bibitem{Sekercioglu08}   
 C.H. Sekercioglu,   Quantifying coauthor contributions,  Science 322 (2008) 371.
 
\bibitem{Schreiber2008d} M. Schreiber,  To share the fame in a fair way, $h_m$ for multi-authored manuscripts,  New J. Phys.  10 (2008) 040201.  

\bibitem{Schreiber2008c} M. Schreiber,    A modification of the h-index: The h(m)-index accounts for multi-authored manuscripts,  J. Informetrics  2 (2008)  211-216. 

 \bibitem{Egghe2008d} L. Egghe, Mathematical theory of the  $h-$ and $g$-index in case of fractional counting of authorship, J.    Am. Soc. Inf. Sci. Technol.    59 (2008) 1608-1616.
 
 \bibitem{Hagen09}      N.T. Hagen,   Credit for coauthors,  Science 323 (2009) 583. 
 
\bibitem{ZhangEMBO10.09} C.T. Zhang, A proposal for calculating weighted citations based on author rank.   EMBO Rep.  10 (2009) 416-417. 


\bibitem{JoI4.10.42Schreibergm}   M. Schreiber,    How to modify the g-index for multi-authored manuscripts,  J.  Informetrics  4 (2010) 42-54. 

\bibitem{Carbone} V. Carbone, Fractional counting of authorship to quantify scientific research output, arxiv 1106.0114v1     (2011). 

 \bibitem{galamscientom12}   S. Galam,  Tailor based allocations for multiple authorship: a fractional gh-index, Scientometrics 89   (2011) 365-379.
 
\bibitem{Sofia3}    
M. Ausloos,   A scientometrics law about co-authors and their ranking. The co-author, Scientometrics 95 (2013) 895-909.
 
\bibitem{[HB]}  
 H. Bougrine,   Subfield Effects on the Core of Coauthors,      Scientometrics 98 (2014) 1047-1064.
 
    \bibitem{[JM]}  
 J. Miskiewicz,   Effects of Publications in Proceedings  on the Measure of the Core Size of Coauthors.   Physica A 392 (2013) 5119-5131.
 
        \bibitem{[GR]}   
G. Rotundo, Black-Scholes-Schrodinger-Zipf-Mandelbrot model framework for improving a study of the coauthor core score, Physica A 404 (2014)  296-301. 

    \bibitem{profitindexPLOS} N.A. Aziz,  M.P. Rozing, 
 Profit (p)-index: the degree to which authors profit from co-authors,   PloS One  8 (2013) e59814.
    
     \bibitem{Kwok05}  L.S. Kwok,  The White Bull effect: abusive coauthorship and publication parasitism, J. Med. Ethics  31 (2005) 554-556. 

 \bibitem{Morris2005}  S. A. Morris, Manifestation of emerging specialties in journal literature: A growth model of papers, references, exemplars, bibliographic coupling, co-citation, and clustering coefficient distribution,  J.    Am. Soc. Inf. Sci. Technol.    56  (2005) 1250-1273.
 
 \bibitem{JASIST48.07.1764coauth}   S. A. Morris,
    M. L. Goldstein, Manifestation of research teams in journal literature: A growth model of papers, authors, collaboration, coauthorship, weak ties, and Lotka's law,  J.    Am. Soc. Inf. Sci. Technol.    58   (2007) 1764-1782.
 
  \bibitem{sidiropoulosscientom10} A. Sidiropoulos, D.  Katsaros, Y. Manolopoulos, Generalized Hirsch h-index for disclosing latent facts in citation networks, Scientometrics 72 (2007) 253-280
  
    \bibitem{JoI7.13.198AbramoorderCA}  G. Abramo,  C. A. D'Angelo,   F. Rosati,
 The importance of accounting for the number of co-authors and their order when assessing research performance at the individual level in the life sciences,  J.  Informetrics  7 (2013) 198--208. 
 

\end{thebibliography}
 \end{document}